\documentclass[10pt,a4paper]{article}
\usepackage{amsfonts}
\usepackage{graphicx}
\title{U(1) Gauge Field in 6D Space-Time With Compact Noncommutative Dimensions:\\
A Coherent State Approach}
\author{M. $\textrm{Nasseri}^1, \textrm{A. Jahan}^2$, M. Souri\\
${}^{1}\textrm{Isalmic}$ Azad Univeristy of Hashtgerd, Hashtgerd, Tehran, Iran\\
${}^{2}\textrm{RIAAM}$, P. O. Box: 55134 - 441, Maragha, Iran \\\ jahan@riaam.ac.ir.}
\date{}
\usepackage[english]{babel}
\begin{document}
\maketitle
\begin{abstract}
We consider the U(1) gauge field defined over a six dimensional space-time with extra dimensions compactified on a noncommutative toroidal orbifold, within the context of coherent state approach to the noncommutative spaces. We demonstrate that the fuzzines of extra dimensions can lead to the canceling of the part of electrostatic interaction mediated by the massive KK modes.
\end{abstract}
\section{Introduction}
Quantum theory of fields on noncommutative space-times, recognized as noncommutative field theories, generalizes the familiar notions of
the usual field theory to the case of noncommutative space-time. Although, the idea of noncommutating space-time coordinates is
an old proposal [1], but the recent discoveries in string/M theories were the main source of
renewed interests in the subject [2]. In noncommutative field theories
the space-time coordinates $x^{\mu}$ are replaced with noncommutating coordinates $\hat{x}^{\mu}$, satisfying
\begin{equation}\label{1}
[\hat{x}^{\mu},\hat{x}^{\nu}]=i\theta^{\mu\nu},
\end{equation}
where $\theta^{\mu\nu}$   stands for a real antisymmetric matrix. Thus because of the uncertainty relation
\begin{equation}\label{2}
\Delta{x^{\mu}}\Delta{x^{\nu}}\geq{\frac{\theta^{\mu\nu}}{2}},
\end{equation}
induced by the equation (1), there will be a mixing between the short and large distance scales implying for a mixing between the ultraviolet and infrared behaviors of the field theories in noncommutative
 space-times. Such a problem, which is known as the "UV/IR mixing problem", is a distinct feature of the noncommutative models [3].
In a noncommutative space-time the usual product between the fields must be replaced by the Weyl-Moyal or  star-product
defined as
{\setlength\arraycolsep{2pt}
\begin{eqnarray}\label{3}
\big(f\star{}g\big)(x)&=&\lim_{x\to{y}}e^{\frac{i}{2}\theta^{\mu\nu}\partial^{x}_\mu\partial^{y}_\nu}f(x)g(y)\\
&=&\lim_{x\to{y}}f(x)g(y)+\frac{i}{2}\theta^{\mu\nu}\partial^{x}_\mu\partial^{y}_\nu{}f(x)g(y)
+O(\theta^{2}).\nonumber
\end{eqnarray}}
So, because of the infinite series of the derivatives, star-product reveals an inherent nonlocality accompanying the theory.\\
Recently, a new approach with a completely different point of view is proposed to study the noncommutative space-times [4]. The main idea is to use the expectation values of the operators between the coherent states of the noncommutative space-time instead of using the star-product (3).
In this approach the free particle propagator acquires a Gaussian damping factor, which may act as a probable mechanism to make the whole
formalism of the perturbation theory finite. So in the coherent state approach the noncommutativity of space-time manifests itself by
modification of the propagators rather than the interaction vertices and this is to say that there is no UV/IR mixing problem in coherent
state approach to the noncommutative space-time. On the other hand, recently some models of the noncommutative space-times proposed by
some authors in which the noncommutativity is restricted to the compact extra dimensions, leaving the four non-compact
dimensions commutative [5, 6] (see also [7] for a pedagogical introduction to extra dimensions). For example, the model developed in [6] assumes space-time with noncommutative extra dimensions
compactified on a toroidal orbifold $T^{2}/\mathbb{Z}_{2}$. The massless gauge filed is defined over the six dimensional bulk while
the matter field is confined to the four non-compact dimensions. The effective theory in four dimensions consists of massless
photons and massive Kaluza-Klein (KK) modes. In this letter we shall study the effect of
noncommutativity of a toroidal orbifold on the electrostatic potential within the context of the coherent state approach to
the noncommutative space. We show that in this approach the fuzziness of the compact extra dimensions acts as a cut-off over the KK modes and thus leads to cancelation of the contributions made by the massive KK modes in the $\theta\rightarrow\infty$.
\section{Coherent State Approach to the Noncommutative Space}
Let's consider a  2d dimensional noncommutative space-time. It is always possible to find a set of coordinates such that [4]
\begin{equation}\label{4}
[\hat{y}^{2}_{i},\hat{y}^{1}_{i}]=i\theta,
\end{equation}
where $\vec{y}_{i}=(y^{2}_{i},y^{1}_{i})$ which is the position vector of the i-th plane. We also assume for the two distinct points $\vec{y}_{i}$ and $\vec{y}^{\,\prime}_{i}$
\begin{equation}\label{4}
[\hat{y}^{\prime{}2}_{i},\hat{y}^{2}_{i}]=[\hat{y}^{\prime1}_{i},\hat{y}^{1}_{i}]=[\hat{y}^{\prime2}_{i},\hat{y}^{1}_{i}]=0.
\end{equation}
The associated raising and lowering operators are
{\setlength\arraycolsep{2pt}
\begin{eqnarray}\label{5}
A_{i}=\frac{1}{\sqrt{2}}(\hat{y}^{2}_{i}+i\hat{y}^{1}_{i}),\\
A^{\dag}_{i}=\frac{1}{\sqrt{2}}(\hat{y}^{2}_{i}-i\hat{y}^{1}_{i}).\nonumber
\end{eqnarray}}
Coherent states corresponding to the above operators are defined as the eigenstates $|\alpha_{i}\rangle$ in the following sense
{\setlength\arraycolsep{2pt}
\begin{eqnarray}\label{6}
A_{i}|\alpha_{i}\rangle&=&\alpha_{i}|\alpha_{i}\rangle,\\\langle\alpha_{i}|A^{\dag}_{i}&=&\langle\alpha_{i}|\alpha^{*}_{i}.\nonumber
\end{eqnarray}}
Then the mean position of the particle on the i-th plane is defined as
\begin{eqnarray}\label{7}
\vec{x}_{i}=\langle\alpha_{i}|\vec{\hat{y}}_{i}|\alpha_{i}\rangle.
\end{eqnarray}
More generally, any function $\hat{F}(\hat{y})$ defined over the noncommutative space-time will be replaced by its mean value as
{\setlength\arraycolsep{2pt}
\begin{eqnarray}\label{8}
F_{\theta}(x)&=&\prod_{k=1}^{d}\langle\alpha_{k}|\hat{F}(\hat{y})|\alpha_{k}\rangle
=\prod_{k=1}^{d}\langle\alpha_{k}|\int\frac{d^{d}\vec{p}_{k}}{(2\pi)^d}f(p)e^{i\vec{p}_{k}\cdot\vec{\hat{y}_{k}}}|\alpha_{k}\rangle\\
&=&\prod_{k=1}^{d}\int\frac{d^{d}\vec{p}_{k}}{(2\pi)^d}f(p)e^{i\vec{p}_{k}\cdot\vec{x}_{k}}e^{-\frac{\theta}{2}\vec{p}^{2}_{k}}\nonumber\\
&=&\int\frac{d^{\,2d}p}{(2\pi)^{2d}}f(p)e^{ipx}e^{-\frac{\theta}{2}p^{2}},\quad\quad\quad\quad\quad\vec{p}_{k}=(p^{2}_{k}, p^{1}_{k}).\nonumber
\end{eqnarray}}
where $p^{\mu}=(\vec{p}_{1},\ldots,\vec{p}_{k})$ and $p^{2}=p^{\mu}p_{\mu}=\sum^{d}_{k=1}\vec{p}^{\,2}_{k}$. So, in coherent state approach the fuzziness of space-time manifests itself by smearing the fields over space by modification of the kernel
of integral via a Gaussian damping factor as in (9).  In particular for the Euclidean free particle propagator we get
\begin{equation}\label{9}
g(p)=\frac{e^{-\frac{\theta}{2}p^{2}}}{p^{2}+m^{2}}.
\end{equation}
\section{U(1) Gauge Field on $T^{2}/\mathbb{Z}_{2}$ Orbifold}
In this section we briefly discuss the model considered in [6]. The model assumes the U(1) gauge field to be defined over the six dimensional
bulk, while the matter field is confined to the non-compact four dimensions. The two extra dimensions are compactified on a noncommutative
toroidal orbifold. The six dimensional Lagrangian of the free gauge field plus the gauge fixing term is
\begin{equation}\label{10}
\mathcal{L}_{6D}=-\frac{1}{4}\mathcal{F}^{MN}\mathcal{F}_{MN}-\frac{1}{2}\zeta\bigg(\partial_{\mu}
\mathcal{A}^{\mu}+\frac{1}{\zeta}\partial_{m}
\mathcal{A}^{m}\bigg)^{2},\quad{m=4,5}.
\end{equation}
with $X^{M}=(x^{\mu},y^{m})$ as a position 6-vector in six dimensional bulk. Note that we have maintained the usual product between the fields as we are considering the model within the context of coherent state approach [6]. The effective Lagrangian in four dimensions decomposes to the Kaluza-Klein (KK) modes as
{\setlength\arraycolsep{2pt}
\begin{eqnarray}\label{11}
\mathcal{L}_{4D}&=&-\frac{1}{4}F^{(\vec{0})\mu\nu}F^{(\vec{0})}_{\mu\nu}-\frac{1}{2}\zeta\big(\partial_{\mu}
A^{\mu(\vec{0})}\big)^{2}\\
&+&\sum_{\vec{n}\neq{\vec{0}}}\left(-\frac{1}{4}F^{(\vec{n})\mu\nu}F^{(\vec{n})}_{\mu\nu}+m^{2}
_{\vec{n}}A^{\mu(\vec{n})}A_{\mu}^{(\vec{n})}-\frac{1}{2}\zeta\big(\partial_{\mu}
A^{\mu(\vec{n})}\big)^{2}\right) {}\nonumber\\
&+&\sum_{\vec{n}\neq{\vec{0}}}\left(\frac{1}{2}\partial^{\mu}A^{(\vec{n})}_{L}\partial_{\mu}A^{(\vec{n})}_{L}+\frac{1}{2}m^{2}
_{\vec{n}}A_{L}^{(\vec{n})}A_{L}^{(\vec{n})}\right)\nonumber\\
&+&\sum_{\vec{n}\neq{\vec{0}}}\left(\frac{1}{2}\partial^{\mu}A^{(\vec{n})}_{H}\partial_{\mu}A^{(\vec{n})}_{H}+\frac{1}{2\zeta}m^{2}
_{\vec{n}}A_{H}^{(\vec{n})}A_{H}^{(\vec{n})}\right),\nonumber
\end{eqnarray}}
with $\vec{n}=0$ or $\vec{n}=(n_{1}=0, n_{2}>0)$ or $\vec{n}=(n_{1}>0, n_{2}=0,\pm1\ldots)$. Every individual KK mode has the rest mass
\begin{equation}\label{12}
m^{2}_{\vec{n}}=\frac{n^{2}_{1}}{R^{2}_{1}}+\frac{n^{2}_{2}}{R^{2}_{2}}.
\end{equation}
where $R_{1}$   and $R_{2}$   stand for the principal radii of the tours and the fields $A_{L}$ and $A_{H}$ are the physical and un-physical scalars respectively. In the limit $\zeta\to{0}$   ,
the scalar $A^{(\vec{n})}_{H}$   decouples from the Lagrangian (12). So, the ordinary massless photons, massive KK photons and the massive KK scalars are the three physical fields incorporating the free field Lagrangian. The effective interaction term in four dimension is
\begin{equation}\label{13}
\mathcal{L}_{4D, int}=q\bar{\psi}{(x)}\gamma_{\mu}A^{\mu(\vec{0})}(x)\psi{(x)}
+q\sqrt{2}\sum_{\vec{n}\neq{0}}\bar{\psi}{(x)}\gamma_{\mu}A^{\mu(\vec{n})}(x)\psi{(x)}.
\end{equation}
The KK modes satisfy the Lorentz gauge condition, i.e. $\partial_{\mu}A^{\mu(\vec{n})}=0$, automatically. We assume that the massless modes also satisfy
the Lorentsz gauge, i.e. $\partial_{\mu}A^{\mu(\vec{0})}=0$  . Since the noncommutativty restricted to the two extra dimensions, i.e. $[\hat{y}^{5},\hat{y}^{4}]=i\theta$, the free propagator in six dimensions reads
\begin{equation}\label{14}
g_{MN}(k)=\frac{e^{-\frac{\theta}{2}(k^{2}_{4}+k^{2}_{5})}}{k^{2}-k^{2}_{4}-k^{2}_{5}}\eta_{MN}.
\end{equation}
Compactification of the extra dimensions on a torus leads to the quantization of the conjugate momenta, i.e. $(k_{4},k_{5})\to{(\frac{n_{1}}{R_{1}},\frac{n_{2}}{R_{2}})}$. Thus from (12) and (13), for the massless and massive modes we find
{\setlength\arraycolsep{2pt}
\begin{eqnarray}\label{15}
g^{(\vec{n})}_{\mu\nu}(k)&=&\frac{e^{-\frac{\theta}{2}m^{2}_{\vec{n}}}}{k^{2}-m^{2}_{\vec{n}}}\eta_{\mu\nu},\\
g^{(\vec{0})}_{\mu\nu}(k)&=&\frac{1}{k^{2}}\eta_{\mu\nu}.
\end{eqnarray}}
where we have assumed the noncompact dimensions to be Lorentzian. Therefore, the ordinary massless mode is not affected by the noncommutativity of the extra dimensions, but the massive modes get a Gaussian damping factor. Hence, for a time-independent point-like distribution of charge, located at origin, from (16) and (17) one immediately finds
{\setlength\arraycolsep{2pt}
\begin{eqnarray}\label{16}
A^{(\vec{0})}_{\,0}(\vec{x})&=&q\int{\frac{d^{3}k}{(2\pi)^3}}\frac{e^{i\vec{k}\cdot{}\vec{x}}}{\vec{k}\,^2}
=\frac{q}{4\pi{}r},\\
A^{(\vec{n})}_{\,0}(\vec{x})&=&q\sqrt{2}\int{\frac{d^{3}k}{(2\pi)^3}}\frac{e^{i\vec{k}\cdot{}\vec{x}
-\frac{\theta}{2}m^{2}_{\vec{n}}}}{\vec{k}\,^2+m^{2}_{\vec{n}}}\\
&=&\frac{q\sqrt{2}}{4\pi{r}}\,\frac{1}{\sqrt{\pi}}\int_{0}^{\infty}\frac{ds}{\sqrt{s}}e^{-s}e^{-(\frac{r^{2}}{4s}+\frac{\theta}{2})m^{2}
_{\vec{n}}},\nonumber
\end{eqnarray}}
with $j^{\mu}(x)=\big(\rho(\vec{x}\,),\vec{0}\,\big)=\big(q\delta(\vec{x}\,),\vec{0}\,\big)$ and $r=|\vec{x}|$. So the potential energy of  two particles with charges $q_{1}$ and $q_{2}$ will be
{\setlength\arraycolsep{2pt}
\begin{eqnarray}\label{17}
U_{\theta}(r)&=&\frac{q_{2}q_{1}}{4\pi{}r}\bigg(1+\frac{2}{\sqrt{\pi}}\sum^{\infty}_{\vec{n}\neq{0}}
\int_{0}^{\infty}\frac{ds}{\sqrt{s}}e^{-s}e^{-(\frac{r^{2}}{4s}+\frac{\theta}{2})m^{2}
_{\vec{n}}}\bigg).
\end{eqnarray}}
Then by means of $\sigma_{1,2}(s)=(\frac{r^{2}}{2s}+\theta)/2R_{1,2}$, the expression (20) for the potential energy can be re-expressed as
\begin{equation}\label{19}
U_{\theta}(r)=\frac{q_{2}q_{1}}{4\pi{}r}\frac{1}{\sqrt{\pi}}
\int_{0}^{\infty}\frac{ds}{\sqrt{s}}e^{-s}\Theta(i\sigma_{1}{(s)}/\pi)\Theta(i\sigma_{2}{(s)}/\pi),
\end{equation}
where the Jacobi theta function is defined as
\begin{equation}\label{20}
\Theta(\nu,\tau)=\sum_{n=-\infty}^{+\infty}e^{i\pi{n^{2}}\tau+2i\pi{n}\nu},
\end{equation}
in particular
\begin{equation}\label{21}
\Theta(i\sigma/\pi)\equiv\Theta(0,i\sigma/\pi)=\sum_{n=-\infty}^{+\infty}e^{-\sigma{n^{2}}},
\end{equation}
It should be noticed that in the lack of noncommutativity parameter in Eq. (21), the Jacobi theta function diverges at the upper limit of integration. However, the situation changes when the $\theta$ parameter appears because of the noncommutative nature of extra dimensions. Now, lets define $\rho_{1,2}=r/R_{1,2}$ and $\vartheta_{1,2}=\theta/2R^{2}_{1,2}$. Near the threshold, i.e. $1\sim\rho_{1,2}$, it will be enough to consider just only modes with lowest KK mass [8]. Thus
\begin{equation}\label{22}
U_{\theta}(r)=\frac{q_{1}q_{2}}{4\pi{}r}\bigg[1+2\bigg(e^{-(\vartheta_{1}+r/R_{1})}+e^{-(\vartheta_{2}+r/R_{2})}\bigg)\bigg],
\end{equation}
where we have invoked [9]
\begin{equation}\label{23}
\int_{0}^{\infty}\frac{ds}{\sqrt{s}}\,
e^{-(s+\frac{\alpha}{s})}=\sqrt{\pi}e^{-2\sqrt{\alpha}}.
\end{equation}
However, in the limit $\vartheta_{1,2}\to{}\infty$, second term of (24) vanishes and one may argue that the fuzziness of compact dimensions blocks out the contribution of the higher KK modes to
the potential energy. A rather simple expression for the potential energy can be
achieved, if one integrates over the \textit{s} variable in (20). The result is
\begin{equation}\label{17}
U_{\theta}(r)=\frac{q_{2}q_{1}}{4\pi{}r}\bigg(1+2\sum^{\infty}_{\vec{n}\neq{0}}e^{-m_{\vec{n}}r-\frac{\theta}{2}m^{2}_{\vec{n}}}
\bigg).
\end{equation}
which coincides with (24) when only the modes with lowest KK mass, i.e. $\vec{n}=\{(1,0),(0,1)\}$ is considered.
\section{Conclusion}
Coherent state approach to the compact noncommutative dimensions, provides a mechanism to cancel the contribution coming from the higher KK modes when the extra dimensions become extremely fuzzy. So in the case of a U(1) gauge field defined over a six dimensional space-time with two compact noncommutative dimensions, it will be hard to detect the deviation of the electrostatic potential from the Coulomb law if there is a very large noncommutativity between the compact dimensions.
\section{Acknowledgments}
Authors would like to thank Islamic Azad University of Hashtgerd for their financial support.

\end{document}